\documentclass[%
 reprint,
 amsmath,amssymb,
 aps,
]{revtex4-2}

\usepackage{graphicx}
\usepackage{dcolumn}
\usepackage{bm}

\usepackage{lineno}
\usepackage{enumitem}
\usepackage[percent]{overpic}
\usepackage{graphicx}
\usepackage{subcaption}
\usepackage{dblfloatfix}    
\usepackage[colorlinks=true,allcolors=blue, urlcolor=red]{hyperref}%

\begin{document}

\title{Adaptive Optics-Enhanced Michelson Interferometer\\for Spectroscopy of Narrow-Band Light Sources}

\author{Jesneil Lauren Lewis}
 \email{JesneilLewis@proton.me}
 \altaffiliation{}
\author{Ayan Banerjee}%
 \email{ayan@iiserkol.ac.in}
\affiliation{Department of Physical Sciences (DPS)\\Center of Excellence in Space Sciences India (CESSI)\\Indian Institute of Science Education and Research Kolkata (IISER-K)
}%

\date{\today}

\begin{abstract}
Adaptive optics enables the deployment of interferometer-based spectroscopy without the need for moving parts necessary for scanning the interferometer arms. 
Here, we employ a Michelson Interferometer in conjunction with a Spatial Light Modulator (SLM) for determining the spectral profile of a narrow-band light source. Interestingly, we observe that the fringes across the interferometer output beam are inherently shifted in wavelength even when a constant phase profile is provided to the SLM. We calibrate the spectral shifts as a function of fringe spatial location by measuring the incident light spectrum at various points across the fringe pattern, and observe that the spectral peak traces out a `teardrop' shape, whose width is dependent on the spectral bandwidth of the source, the relative tilt and path difference between the two arms of the interferometer, and the divergence of the beam. Next, we demonstrate that this inherent spectral variation of the fringes can be used to perform fast single-snapshot spectroscopy of narrow-band light sources, while a time-varied phase profile provided to the SLM leads to multi-step spectroscopy with lower noise, higher resolution, and better contrast. Our findings establish that the Michelson Interferometer can be used to perform spectroscopy of any source within a certain spectral range from simple images of the fringe pattern, so as to facilitate exciting applications towards hyperspectral imaging.
\end{abstract}


\maketitle

High-resolution spectroscopy has enormous applications in diverse areas of research, ranging from molecular fingerprinting \cite{mole1, mole2, mole3} to biosensing \cite{biosense1, biosense2} to astronomy \cite{astro2, astro1}. In astronomy, especially, spectrometers are almost ubiquitous, both aboard spacecrafts, and in ground-based stations \cite{space_based, ground_based}. Now, most high-resolution commercial spectrometers are based on diffraction gratings (Echelle spectrometers) \cite{astro1}. However, a large optical path is necessary to obtain the necessary resolution for grating spectrometers, which often renders them bulky. Interferometers -- both Fabry-Perot \cite{fabry1, fabry2}, and Michelson \cite{MI2, MI1} -- mitigate the optical path issue since the mirror reflectivity is another parameter that can be tuned to obtain high finesse (resolution). However, interferometers need to be scanned along the beam path in order to build a spectrum, and this is typically achieved by piezo-electric transducers or motors \cite{fabry1}. This leads to the requirement of moving parts, which may become an issue in unmanned applications, such as those aboard space missions. Scanning may be achieved by tuning the refractive index along the light path with the help of electro-optic crystals that require rather high voltage and also have calibration issues under remote operation \cite{EO-high}.

In this context, adaptive optics offers a potential solution with its inherent flexibility and remote accessibility \cite{aoastro2}, and has been deployed in interferometry successfully \cite{AO_Intf, AO_Intf2}. 
Traditional use of adaptive optics has mostly been limited to the domain of wavefront correction. However, it should also be possible to use an SLM (Spatial Light Modulator) as a proxy to shift the beam in the propagation direction, thereby scanning different wavelengths. Importantly, though, there is an essential difference in the manner in which SLM-based and traditional mirror-based methods of beam translation can employed for spectroscopy. In traditional systems, the mirror translation leads to beam shift in the longitudinal direction which results in different wavelengths satisfying the interference condition. In an SLM-based interferometer \cite{Chandra:22}, the dynamic phase change in the transverse spread of the reflected/transmitted light couples with the longitudinal propagation of the beam and the spectral information of the source to give fringes that have a consistent transverse spectral variation, which can be tuned/controlled by changing the grey value (GV) provided to the SLM. Careful calibration of GV change and tracking of the fringe in both physical space and spectral domain are required to use this property to perform spectroscopy.

In this paper, we work on this aspect, and provide a rigorous analysis of determining wavelength variation across the fringe profile obtained in the output of a Michelson interferometer based on adaptive optics. We use the configuration used earlier in Ref.~\cite{Chandra:22}, where one of the mirrors of the Michelson Interferometer (MI) was replaced by a reflective phase-only SLM. 
Here, we demonstrate how such a configuration can be actually deployed for spectroscopy, and quantify its performance over two different techniques to perform spectroscopy of a narrow-band source (< 15 nm bandwidth). In both methods, we obtain a one-to-one correspondence between the intensity profile of the interference pattern (obtained using the CCD) and the spectral profile of the light (obtained from the spectrometer), giving us a calibration curve between physical position along the spot and its corresponding wavelength. In the first method, we provide a fixed and uniform GV to the SLM, and measure the spectral variation across a single fringe using the fiber optic spectrometer, while simultaneously taking an image of the interference pattern using a CCD camera. As only a single image file suffices to extract the intensity profile, this single-snapshot method provides the benefit of obtaining the spectrum within a short measurement time, at the cost of reduced contrast and lower resolution. This method merely provides an estimate of the spectrum, with the measured and true spectrum precisely matching only at the fringe maximum and drifting further apart as we move towards the minima on either side. The extent of this deviation is dependent on the size of the fringe relative to the spot size, with a larger/wider fringe (created by a low mirror tilt configuration) providing an accurate measurement of the spectrum over a larger range. Adjusting the SLM phase also sets the wavelength of the fringe maximum which can be important if only a part of a spectrum needs to be measured. In the second technique, we track the spectral peak of the intensity maximum of multiple fringes as they are shifted by varying the phase/path difference in the interferometer through GV change of the SLM. As multiple images need to be obtained (one for every GV change) and the intensity and spectral profiles are independently stitched together from the data at each step, the multi-step method provides the benefit of measuring the spectrum with enhanced contrast and a spectral resolution limited by the GV step length/refractive index change, but at the cost of increased measurement and processing time.
Finally, to perform spectroscopy of an unknown source, we obtain an intensity profile using images of its interference pattern, then apply the calibration curve to convert position data to spectral information, thereby giving us its spectrum. We now go on to describe our experimental setup.

\begin{figure}[ht!]
\centering
\fbox{\includegraphics[height=1.8in]{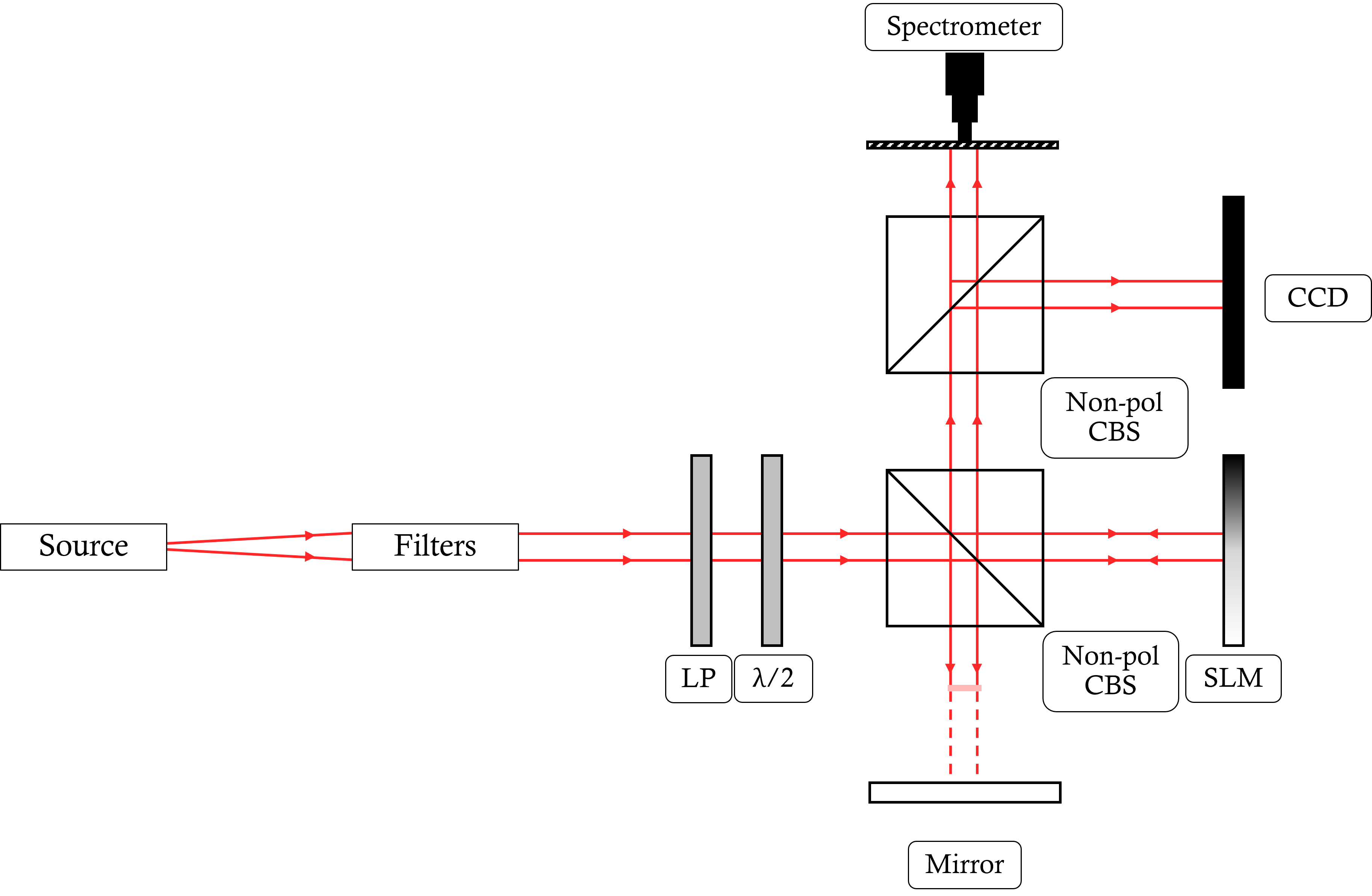}}
\caption{Experimental Setup}
\label{Fig 1}
\end{figure}

A crucial component of the setup (Fig \ref{Fig 1}) consists of various source dependent filters that allow us to pass a collimated narrow-band beam towards the interferometer. The linear polarizer (LP) and $\lambda/2$ waveplate ensure that the polarization axis of this beam is aligned with that of SLM. The beam then passes on from the non-polarizing cube beam splitter (CBS) to the two interferometer arms and back. The mirror is mounted on a translation stage that allows us to set the path difference between the two arms ($\Delta d_{mirror}$) before performing the experiment, with no movement happening afterwards. The beam in contact with the SLM acquires an additional path difference ($\Delta d_{SLM}$) that can varied through GV change, so as to aid in building up the spectrum. Finally, the output beam is passed through another CBS which directs the light to a CCD camera and a fiber optic spectrometer (RI2AS fiber optic spectrometer by Research India), whose head is mounted onto a motorized translation stage. This allows us to obtain the spectrum at any desired point across the wavefront.
This setup enables simultaneous measurement using the spectrometer and the CCD, which is crucial in linking the physical position of the fringe maxima with its corresponding spectral peak, for all GVs. 
Note that in the final implementation of the device, the spectrometer would not be needed post calibration. Indeed, the spectrum will be reconstructed using the intensity profile obtained using the CCD. 

\begin{figure}[ht!]
\centering
\fbox{\includegraphics[height=1.8in]{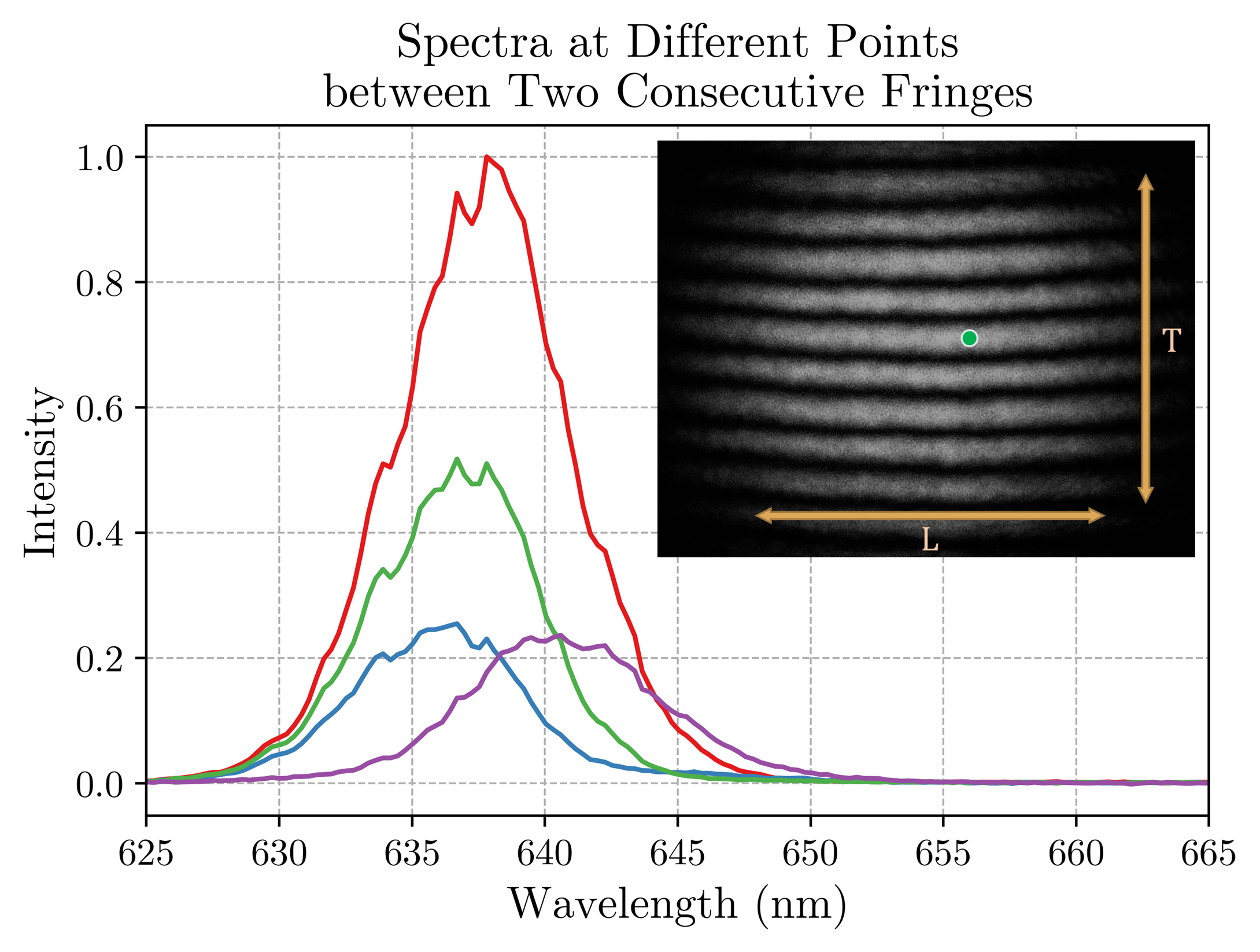}}
\caption{\small \textbf{Subplot} - Fringes with spectrometer position marked in green. Fringes shift in the \textit{T} axis on changing GV. \textbf{Mainplot} - The spectrum recorded at 4 points between the maxima of two consecutive fringes, indicating significant spectral variation.}
\label{Fig 2}
\end{figure}

With this setup, we then sought to measure the nature of spectral variation in the fringe pattern by measuring multiple points in the transverse (\textit{T}) and longitudinal (\textit{L}) directions of the fringe pattern. This can be done by keeping the fringe pattern fixed (set $\Delta d_{mirror}$ and $\Delta d_{SLM}$) and translating the spectrometer over the \textit{T} and \textit{L} axes, or by keeping the spectrometer head fixed and translating the fringe pattern in the \textit{T} axis by changing $\Delta d_{SLM}$. Note, however, that the \textit{L} axis can only be measured using the first method. We observe no spectral variation in the \textit{L} axis (apart from the inherent spatio-spectral variation present for the supercontinuum light source, which is approximately half of the set bandwidth). Using the second method, we see that a 2$\pi$ phase shift in the SLM causes the fringe pattern to shift from one maximum to the next. Thus, a fixed spectrometer head at one fringe maximum scans multiple points between two consecutive fringe maxima. 
Fig \ref{Fig 2} shows the spectrum plotted at four such points, indicating a significant amount of spectral variation in the \textit{T} axis. 

\begin{figure}[ht!]
\centering
\fbox{\includegraphics[height=1.8in]{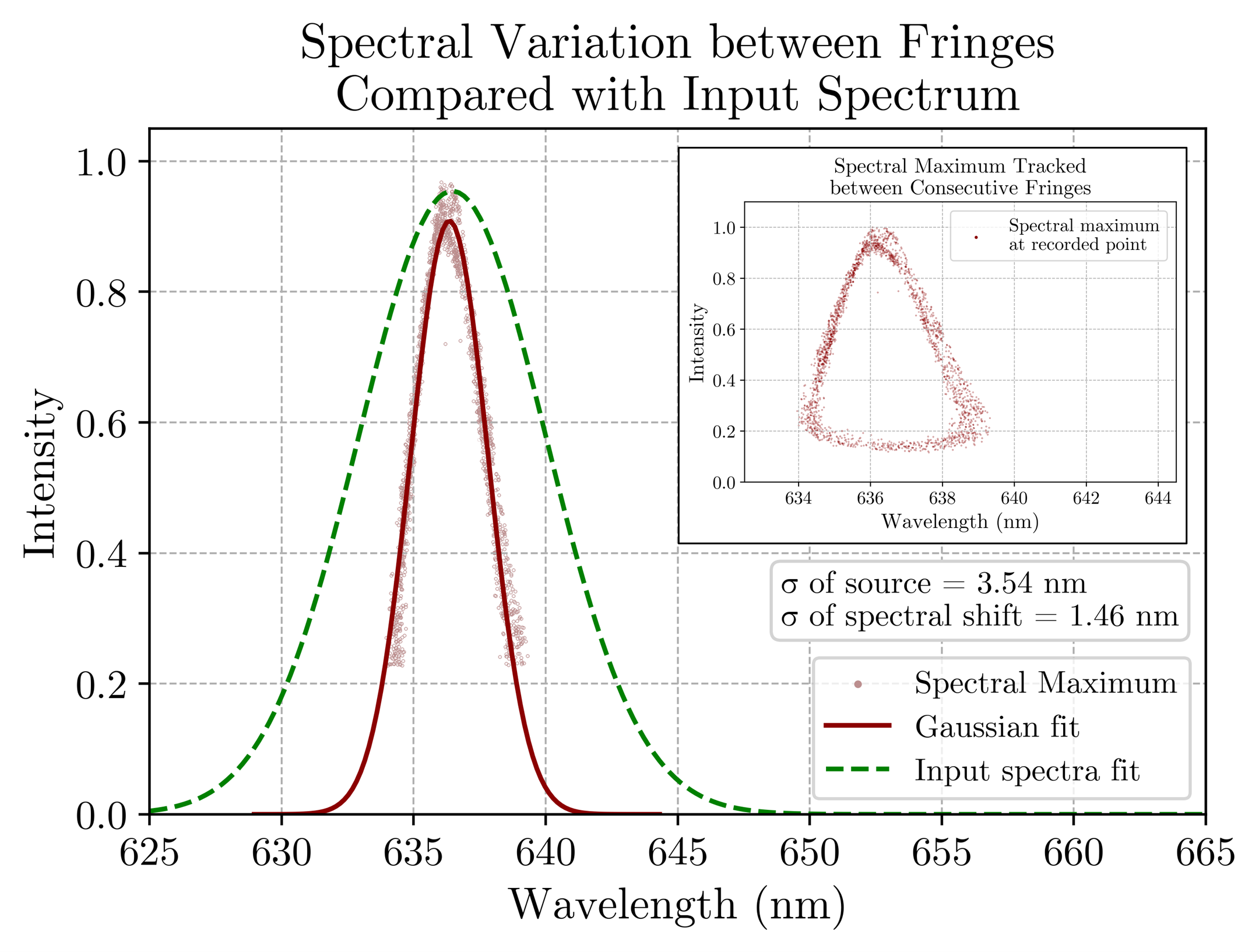}}
\caption{\small \textbf{Subplot} - Spectral maximum tracked across two consecutive fringes. \textbf{Mainplot} - Spectral variation between fringes compared with the spectrum of input light.}
\label{Fig 3}
\end{figure}

Taking a Gaussian fit of the spectrum and tracking the position of the spectral maximum at each recorded point shows us a variation with a `teardrop' shape as seen in the subplot of Fig.~\ref{Fig 3}. The shape is understandable from the fact that the spectrum measurement commences from the fringe maximum which has a spectral peak at $\sim$ 636 nm. On increasing the GV, we observe that the spectrum shifts in an anti-clockwise manner reaching the lower extremum ($\sim$ 634 nm), which is around the half-maximum of the fringe. After this, the spectral peak retraces the wavelengths through the lower portion of the teardrop as it crosses the minima of the fringe pattern, finally shifting to the higher spectral extremum ($\sim$ 639 nm) located around the half-maximum of the adjacent fringe, before returning to the start (which now corresponds to the spectral peak from a point on the next fringe).

To quantify the extent of this spectral variation, we can take a cutoff at intensity $\sim$ 0.2 (which is the threshold below which the spectrum is retraced while crossing across the fringe minima) and perform a Gaussian fit for the rest of the data. The result can be compared to the input spectrum and Fig \ref{Fig 3} shows that the spectral variation is of the same order as the bandwidth of the input light. Further analysis reveals that the extent of the spectral variation between fringes (/$\Delta \lambda$/width of the teardrop) depends on the path difference between the two mirrors. The spectral variation can also be reduced to approximately zero if the path difference is set at the minimization point ($\Delta d_{min}(\lambda)$). Thus, taking $\Delta d = \Delta d_{mirror }+\Delta d_{SLM}$, we have,
\begin{equation}
\Delta \lambda \propto \Delta d ' = \Delta d - \Delta d_{min} 
\label{eq:1}
\end{equation}

\begin{figure}[ht!]
\centering
\fbox{\includegraphics[height=1.5in]{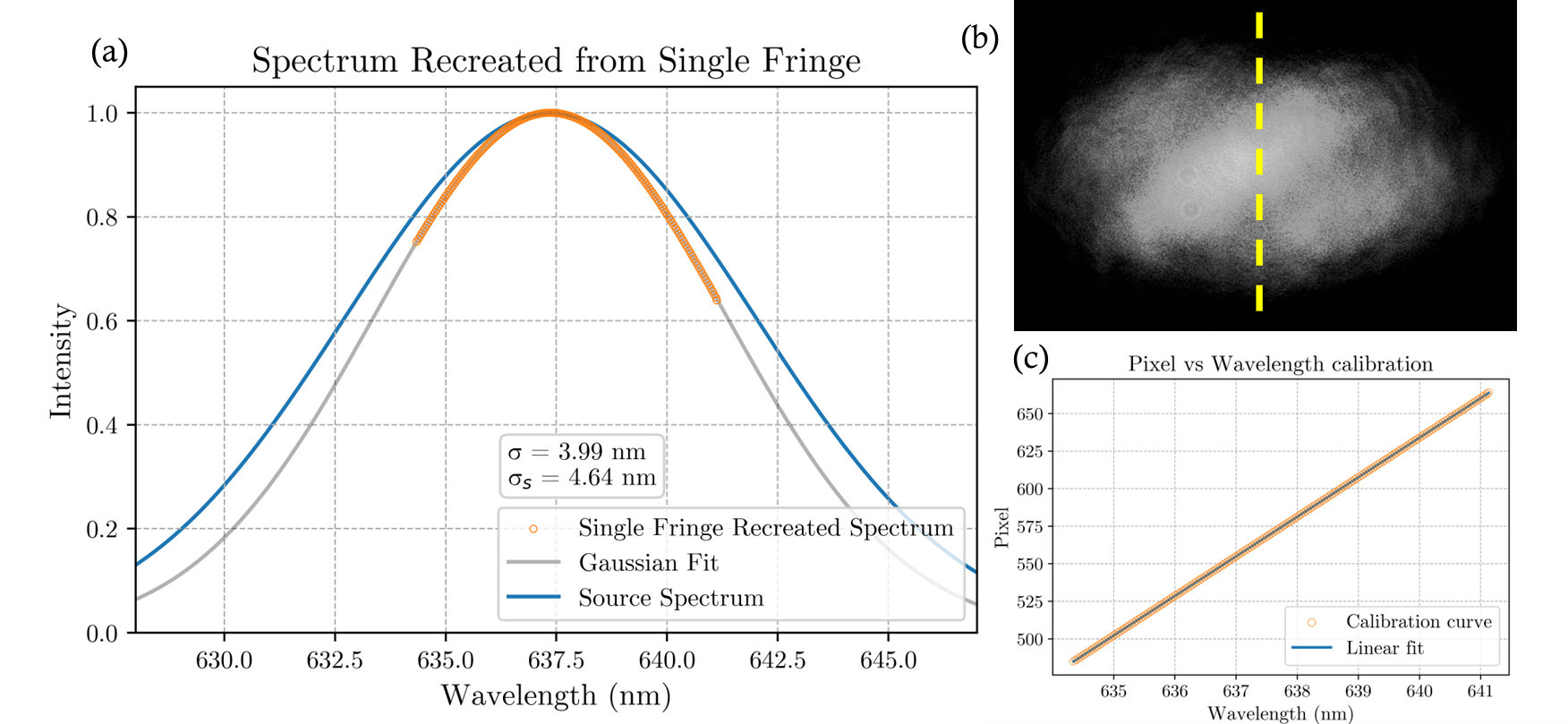}}
\caption{\small \textbf{(a)} Spectrum reconstructed using single-snapshot method compared with true spectrum of source. \textbf{(b)} Large single fringe with spectrometer data recording track marked in yellow. \textbf{(c)} Calibration curve obtained for single-snapshot method}
\label{Fig 4}
\end{figure}

\begin{figure}[ht!]
\centering
\fbox{\includegraphics[height=1.5in]{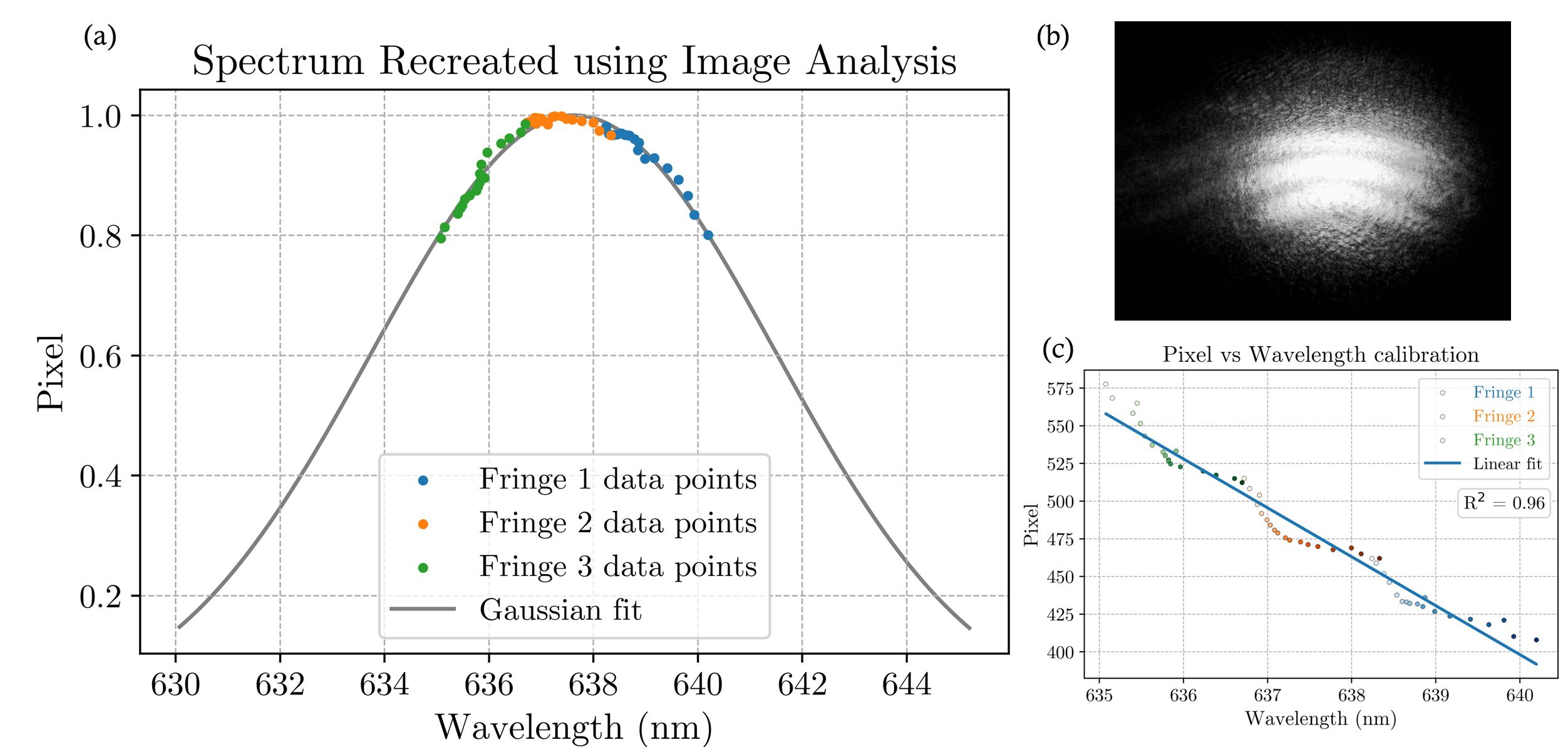}}
\caption{\small \textbf{(a)} Spectrum reconstructed using multi-step method compared with true spectrum of source. \textbf{(b)} Thinner fringes for high-tilt setup. \textbf{(c)} Calibration curve obtained for multi-step method.}
\label{Fig 5}
\end{figure}

Here, $\Delta d'$ is the effective path difference. Note that $\Delta\lambda$ can have both positive and negative values, which manifests in the clockwise and anti-clockwise traversal of the spectral profile in the subplot of Fig ~\ref{Fig 3}, respectively. The length scales of $\Delta d_{mirror}$ and $\Delta d_{SLM}$ are not the same. The translation stage on which the mirror is mounted (providing a translation of $\Delta d_{mirror}$) has a least count of 0.01 mm, while the SLM ($\Delta d_{SLM}$) has a least count of 1 GV, which is $\sim$ 2.48 nm at 635 nm. Thus, the effect of the mirror position ($\Delta d_{mirror}$) has a much greater effect on $\Delta \lambda$. Equation \ref{eq:1} can be expressed as,
\begin{equation}
\Delta \lambda \propto  \Delta d_{mirror} (1+\dfrac{ \Delta d_{SLM}}{\Delta d_{mirror}}) - \Delta d_{min}
\label{eq:2}
\end{equation}
If the mirror is set close to the minimization point ($\Delta d_{mirror} \approx \Delta d_{min}$, ie. $|\Delta d'| \approx 0$), the $\Delta d_{SLM} /\Delta d_{mirror}$ term is significant. Here, $\Delta \lambda$ varies significantly over the GV change range. If the mirror is set away from the minimization point ($|\Delta d'| > 0$), the effect of $\Delta d_{SLM}$ is negligible and $\Delta \lambda$ is maximized, approaching the spectral width of the source. The second case is particularly useful while performing spectroscopy, but reduces the contrast of the fringes for sources with higher bandwidths. The total spectral variation ($\Delta \lambda_{Total}$) can be approximated as a sum of the inter-fringe spectral variations (though there is slight overlap between fringes -- even the teardrop in Fig \ref{Fig 3} includes points neighboring the fringe maxima) and is defined over the beam spot (this property can be seen by comparing Fig \hyperref[Fig 4]{4a} and Fig \hyperref[Fig 5]{5a}). Thus, on increasing the relative tilt between the interfering beams, the beam spot is divided into a greater number of thinner fringes, each having lower $\Delta \lambda$ ($\Delta \lambda_{Total}$ remains fixed), oriented perpendicular to the tilting direction. 
It is also observed that on increasing the divergence of the beam, the $\Delta \lambda$ for a fixed path difference increases. Thus, the factors governing spectral variation are spectral width of the source, the beam divergence, effective path difference, and the relative tilt between interfering beams (this particularly affects the inter-fringe spectral variation -- $\Delta \lambda$). It must be noted that the spectral bandwidth of the source cannot be arbitrarily increased in an attempt to maximize spectral variation of the interference pattern. For larger bandwidths (>15 nm) it has been observed \cite{Chandra:22} that a single point on the intensity profile will have multiple spectral peaks. This would break the unique one-to-one correspondence between points from the intensity and spectral domains required to perform spectroscopy using the methods listed in this paper.

We now move on to a detailed description of the process and results of performing spectroscopy on both single-snapshot and multi-step fronts.
The single-snapshot method starts by applying a fixed and uniform GV to the SLM. We then adjust the relative tilt between the interfering beams such that a single fringe covers the entire beam spot, with both minima adjacent to it clearly detectable. This tilt adjustment is done by physically moving one of the mirrors, but the same effect can be achieved by using different SLM patterns. 
The fringes in Fig \hyperref[Fig 4]{4b} are oriented along a diagonal as this allows for maximum data collection between fringes (length along diagonal is longer than along transverse axis). The spectrometer scans multiple points on the target fringe as it translates along the axis marked in yellow, while a CCD camera captures an image of the interference pattern. Next, the spectral maximum at each recorded point is tracked. Combining the Intensity-Wavelength information from the spectrometer and the Intensity-Position data at each pixel, we get a Position-Wavelength calibration curve(Fig \hyperref[Fig 4]{4c}). This can then be used to find the spectrum of an unknown source by simply using the calibration curve to convert the intensity profile from the CCD image into a spectral profile (shown for a known source in Fig \hyperref[Fig 4]{4a}). The choice of a particular GV on the SLM changes the value of the spectral maximum recorded, and acts as an overall spectral offset.

The multi-step measurement technique is performed at a higher tilt configuration, and requires three key steps. First, we provide a uniform GV to the SLM, but now, increase this GV incrementally from 0 to 255. Meanwhile, we simultaneously measure the interference pattern with the spectrometer (along \textit{T} axis) and the CCD, at each GV step. Next, we track the physical position and spectral peak of each fringe maxima at each GV using the Intensity-Position (intensity profile of interference pattern) data from the CCD and Intensity-Wavelength (spectral profile) information from the spectrometer. Thus, data from each GV is stitched together to obtain the complete intensity and spectral profile. Lastly, we combine the two data sets to obtain a Position-Wavelength calibration curve (refer to Fig \hyperref[Fig 5]{5c}). Once again, this calibration curve can be used to find the spectrum of an unknown source by converting the intensity profile data obtained from images captured using the CCD into a spectral profile of the source. Note that the lighter points in the calibration curve indicate lower GV while darker points indicate higher GV (to help distinguish the direction of spectral shift in fringes.)

\begin{figure}[ht!]
\centering
\fbox{\includegraphics[height=1.7in]{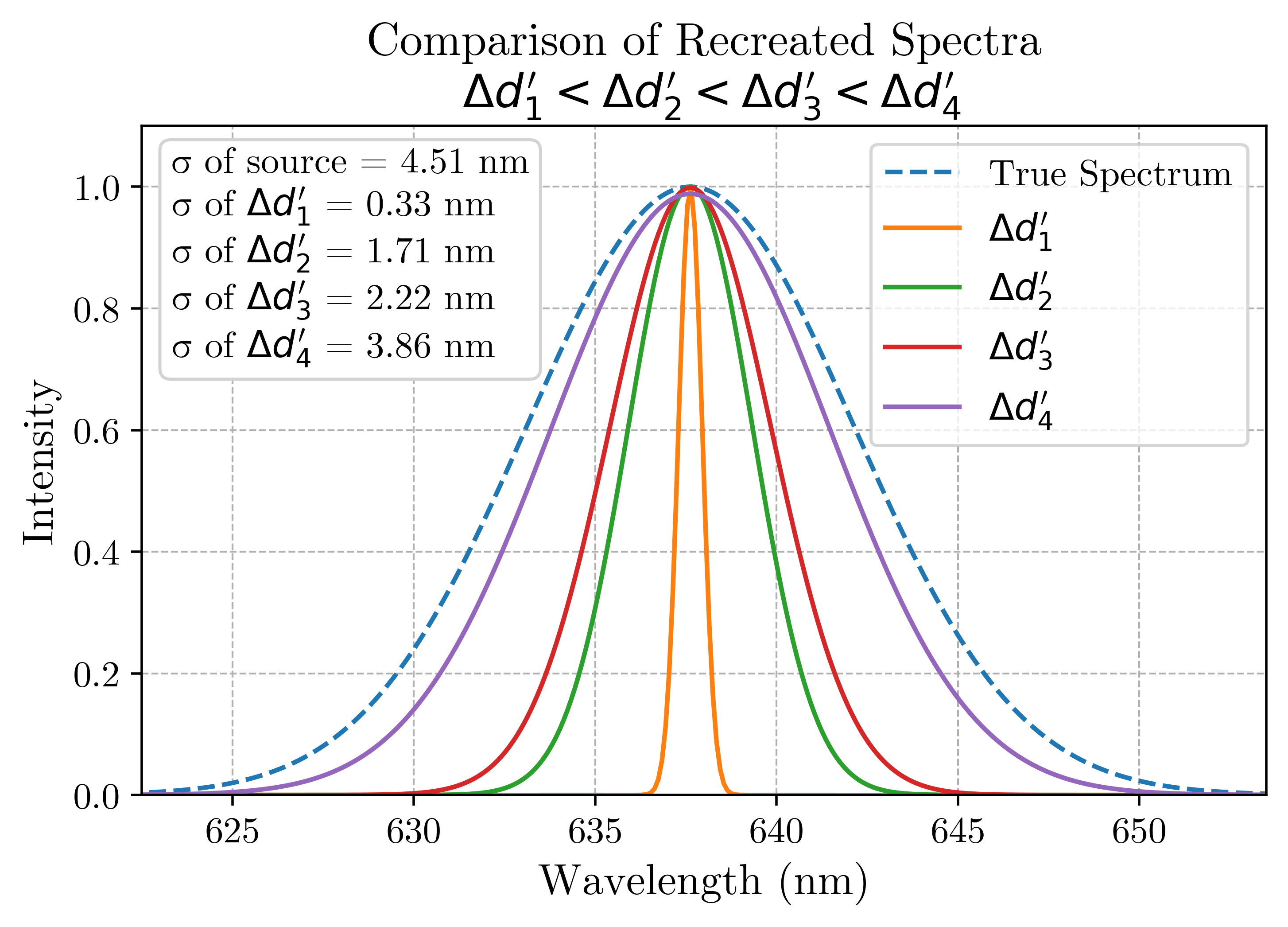}}
\caption{\small Recreated spectra for supercontinuum source with $\Delta d'_1 <\Delta d'_2 <\Delta d'_3 <\Delta d'_ 4$, compared with the spectrum of source}
\label{Fig 6}
\end{figure}

Fig \ref{Fig 6} shows how the effective path difference affects the ability to recreate the spectrum, by comparing them with the true spectrum (outermost (blue) curve). We observer larger spectral variation with higher effective path difference. However, this comes at the cost of lower fringe contrast. In Fig \hyperref[Fig 4]{4a} and Fig \hyperref[Fig 5]{5a}, we see that the spectral shift of each fringe maximum occupies a different portion of the spectrum. These regions can overlap if the phase shift exceeds 2$\pi$ . 

\begin{figure}[ht!]
\centering
\fbox{\includegraphics[height=1.5in]{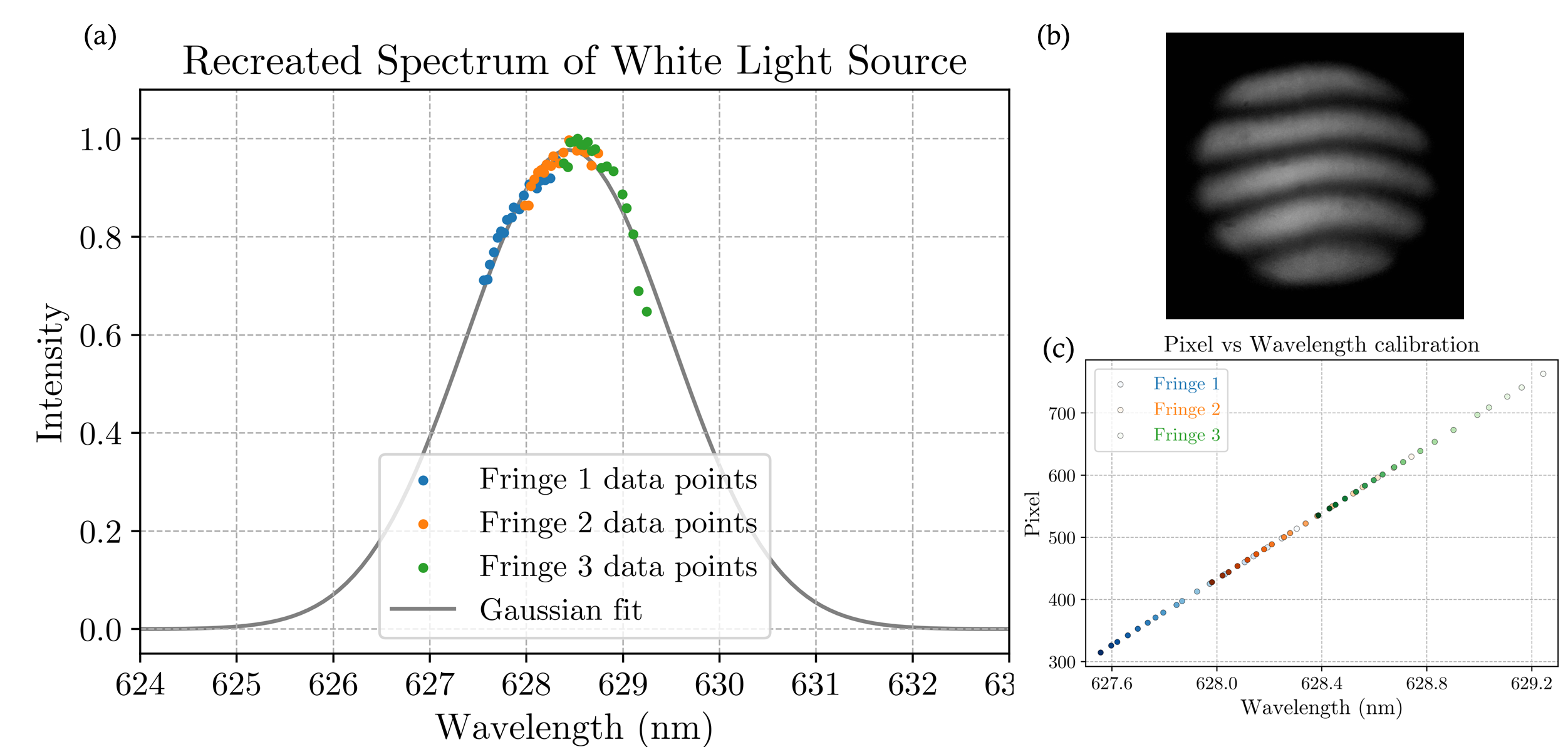}}
\caption{\small \textbf{(a)} Spectrum reconstructed for one of the spectral peaks of a Mercury lamp source at low $\Delta d'$ configuration, leading to lower spectral width. \textbf{(b)} Thinner fringes for high-tilt setup. \textbf{(c)} Calibration curve obtained.}
\label{Fig 7}
\end{figure}

To test whether these properties hold true for different sources, we employed the multi-step spectroscopy method on a Mercury lamp source and successfully generated a calibration curve for the same (Fig \hyperref[Fig 7]{7a} and \hyperref[Fig 7]{7c}). The measurements taken were at a lower $\Delta d'$ and hence the spectral width is lower than the true width of the source, which is in line with the results of Fig \ref{Fig 6}. This confirms the formation of a unique calibration curve over a specific wavelength range. Thus, the use of filters is imperative as it sets the calibration range, and spectroscopy cannot be achieved on sources below the bandwidth of the filter.

In conclusion, we demonstrate two independent methodologies for using an adaptive optics-based MI for spectroscopy of narrow-band sources. We first map out the spectral variation within the MI's fringe pattern. This also reveals the parameters that govern this spectral variation, namely, effective path difference, relative tilt/fringe thickness, beam divergence, and source bandwidth. This spectral variation is then used to perform single-snapshot or multi-step spectroscopy. The single-snapshot method uses a single fringe and tracks the spectral maximum of various points between two consecutive fringe minima. The multi-step method requires tracking the spectral maximum of fringe maxima at each GV. Both methods require simultaneous measurement by a spectrometer and a CCD for calibration. Combining the Intensity-Wavelength and Intensity-Position data sets produces a unique Position-Wavelength calibration for a particular wavelength range. Spectroscopy can now be performed using an intensity profile extracted from images of the interference pattern, with the calibration curve. Note that, if only a small portion of the spectrum needs to be measured with high cadence, and a lower spectral resolution suffices, then the single-snapshot method together with a low tilt configuration would be better, whereas, if an accurate, broad, and detailed spectrum is required, the multi-step method with a higher tilt arrangement would be ideal. We also observe drifts in the fringes over large time scales of the order of several minutes, which may be due to temperature or laser fluctuations - however, this is rather negligible for shorter time scales, which is clear from the fact that we obtain stable data without locking the interferometer path length. Further testing is required to verify if this method of spectroscopy yields similar results for fine and hyperfine atomic spectra, sources with non-Gaussian spectra, and Doppler shifted sources. Needless to mention, stabilizing the length and temperature control of the setup would drastically reduce drifts. We are presently working in these areas and hope to report interesting results soon.

\bibliography{citations}

\end{document}